\documentclass{aa}

\usepackage{graphicx}
\usepackage{txfonts}
\usepackage{natbib}
\usepackage{array}
\usepackage{xspace}
\usepackage{lscape}
\usepackage{url}
\usepackage{arydshln}
\usepackage{rotating}

\newcommand{\kms}{\ensuremath{\mathrm{km\ s^{-1}}}\xspace}

\newcommand{\rrc}{{\it RRC}\xspace}
\newcommand{\vgau}{{\it vgau}\xspace}
\newcommand{\gaussian}{{\it Gaussian}\xspace}

\newcommand{\file}{{\it file}\xspace}
\newcommand{\xmm}{{\it XMM-Newton}\xspace}

\newcommand{\chandra}{{\it Chandra}\xspace}

\newcommand{\Fuse}{{FUSE}\xspace}

\begin{document}

\title{Anatomy of the AGN in NGC 5548}
\subtitle{V. A Clear view of the X-ray Narrow Emission Lines}

\author{M. Whewell \inst{1} 
\and G. Branduardi-Raymont \inst{1}
\and J.S. Kaastra \inst{2, 3, 4}
\and M. Mehdipour \inst{2, 1}
\and K.C. Steenbrugge \inst{5}
\and S. Bianchi \inst{6}
\and E. Behar \inst{7}
\and J. Ebrero \inst{2, 8}
\and M. Cappi \inst{9}
\and E. Costantini \inst{2}
\and B. De Marco \inst{10}
\and L. Di Gesu \inst{2}
\and G.A. Kriss \inst{11, 12}
\and S. Paltani \inst{13}
\and B.M. Peterson \inst{14, 15}
\and P.-O. Petrucci \inst{16, 17}
\and C. Pinto \inst{18}
\and G. Ponti \inst{10}
}

\institute{Mullard Space Science Laboratory, University College London, Holmbury St. Mary, Dorking, Surrey, RH5 6NT, UK \\{\email{m.whewell@ucl.ac.uk}} 
\and SRON Netherlands Institute for Space Research, Sorbonnelaan 2, 3584 CA Utrecht, the Netherlands
\and Department of Physics and Astronomy, Universiteit Utrecht, P.O. Box 80000, 3508 TA Utrecht, the Netherlands
\and Leiden Observatory, Leiden University, P.O. Box 9513, 2300 RA Leiden, the Netherlands
\and Instituto de Astronom\'ia, Universidad Cat\'olica del Norte, Avenida Angamos 0610, Casilla 1280, Antofagasta, Chile
\and Dipartimento di Matematica e Fisica, Universit\`a degli Studi Roma Tre, via della Vasca Navale 84, 00146 Roma, Italy
\and Department of Physics, Technion-Israel Institute of Technology, Haifa 32000, Israel
\and European Space Astronomy Centre, P.O. Box 78, E-28691 Villanueva de la Ca\~{n}ada, Madrid, Spain
\and INAF-IASF Bologna, Via Gobetti 101, I-40129 Bologna, Italy
\and Max-Planck-Institut f\"ur extraterrestrische Physik, Giessenbachstrasse, D-85748 Garching, Germany
\and Space Telescope Science Institute, 3700 San Martin Drive, Baltimore, MD 21218, USA
\and Department of Physics and Astronomy, The Johns Hopkins University, Baltimore, MD 21218, USA
\and Department of Astronomy, University of Geneva, 16 Ch. d'Ecogia, 1290 Versoix, Switzerland
\and Department of Astronomy, The Ohio State University, 140 W 18th Avenue, Columbus, OH 43210, USA
\and Center for Cosmology \& AstroParticle Physics, The Ohio State University, 191 West Woodruff Ave., Columbus, OH 43210, USA
\and Univ. Grenoble Alpes, IPAG, F-38000 Grenoble, France
\and CNRS, IPAG, F-38000 Grenoble, France
\and Institute of Astronomy, University of Cambridge, Madingley Road, Cambridge, CB3 0HA, UK 
}

\date{Received 15 June 2015; Accepted 15 July 2015}

\abstract{Our consortium performed an extensive multi-wavelength campaign of the nearby Seyfert 1 galaxy NGC 5548 in 2013-14. The source appeared unusually heavily absorbed in the soft X-rays, and signatures of outflowing absorption were also present in the UV. He-like triplets of neon, oxygen and nitrogen, and radiative recombination continuum (RRC) features were found to dominate the soft X-ray spectrum due to the low continuum flux.}{Here we focus on characterising these narrow emission features using data obtained from the XMM-Newton RGS (770\,ks stacked spectrum).}{We use SPEX for our initial analysis of these features. Self-consistent photoionisation models from Cloudy are then compared with the data to characterise the physical conditions of the emitting region.}{Outflow velocity discrepancies within the O VII triplet lines can be explained if the X-ray narrow-line region (NLR) in NGC 5548 is absorbed by at least one of the six warm absorber components found by previous analyses. The RRCs allow us to directly calculate a temperature of the emitting gas of a few eV ($\sim$10$^{4}$\,K), favouring photoionised conditions. We fit the data with a Cloudy model of log $\xi$ $=$ 1.45 $\pm$ 0.05\,erg\,cm\,s$^{-1}$, log $N_H$ $=$ 22.9 $\pm 0.4$\,cm$^{-2}$ and log v$_{turb}$ $=$ 2.25 $\pm$ 0.5\,\kms for the emitting gas; this is the first time the X-ray NLR gas in this source has been modelled so comprehensively. This allows us to estimate the distance from the central source to the illuminated face of the emitting clouds as $13.9\pm0.6$\,pc, consistent with previous work.}{}

\authorrunning{M. Whewell et al.}
\titlerunning{Anatomy of the AGN in NGC 5548. V}
\maketitle

\section{Introduction}
\label{intro_sect}

\subsection{Recent observational history of NGC 5548}

\object{NGC 5548} is a well studied, nearby ($z = 0.017175$) AGN historically showing typical Seyfert 1 behaviour, with strong broad emission lines and a small number of narrow emission lines across its optical and UV spectra \citep[e.g.,][]{1995ApJS...97..285K,Peterson:2002cg,2000ApJ...528..292C}. In the soft X-ray band there are also broad and narrow emission lines, seen by \chandra LETGS in 2000 \citep{Kaastra:2000uq}.

As presented in \cite{Kaastra:2014ip}, \object{NGC 5548} has recently undergone a drastic change in its observed spectrum, with the soft X-ray flux 25 times lower than the typical median in observations taken by \chandra in 2002. This suppression of the soft X-ray spectrum continued throughout all 14 \xmm observations within the \cite{Kaastra:2014ip} campaign. Six of the 14 \xmm observations were taken simultaneously with UV spectra which display new broad absorption features. The UV spectra also show narrow absorption lines at velocites associated with the classical WAs in this source, but at lower ionisations than previously observed.
\cite{Kaastra:2014ip} conclude this is due to the presence of a new ``obscurer'' located closer to the source than the classical warm absorbers (WA) and narrow-line region (NLR), which must have entered our line of sight between 2007 and 2011.
This ``obscurer'' directly causes the additional absorption in the soft X-ray band and the broad absorption features in the UV. The classical WAs known in this source (in both X-ray and UV) are shielded from the ionising UV/X-ray radiation by the ``obscurer'' so are now at lower ionisations than their historical values. The 2013-14 ionisation parameters for each component of the X-ray WAs are shown in Table \ref{WA_table}.

\begin{table*}
\begin{minipage}[t]{\hsize}
\setlength{\extrarowheight}{3pt}
\caption{Parameters of the 2013-14 warm absorber components (WA comp.) from \cite{Kaastra:2014ip}}
\label{WA_table}
\renewcommand{\footnoterule}{}
\begin{tabular}{l c c c c | l c c c c}
\hline \hline
WA & $\log\xi$ & $N_{H}$ & $v$ & $\sigma_{v}$ & WA & $\log\xi$ & $N_{H}$ & $v$ & $\sigma_{v}$ \\
comp. & ($10^{-9}$ Wm) & ($10^{20}$ cm$^{-2}$) & (\kms) & (\kms) & comp. & ($10^{-9}$ Wm) & ($10^{20}$ cm$^{-2}$) & (\kms) & (\kms) \\  
\hline
A & 0.33 & 2.0$\pm$0.6 & $-588\pm$34 & 210$\pm$40 & D & 1.91 & 10.7$\pm$11.6 & $-254\pm$25 & 68$\pm$14 \\
B & 1.06 & 7.0$\pm$0.9 & $-547\pm$31 & 61$\pm$15 & E & 2.48 & 28$\pm$8 & $-792\pm$25 & 24$\pm$12 \\
C & 1.70 & 15$\pm$3 & $-1148\pm$20 & 19$\pm$6 & F & 2.67 & 57$\pm$17 & $-1221\pm$25 & 34$\pm$13 \\
\hline
\hline
\end{tabular}
\end{minipage}
\end{table*}

This view of the soft X-ray band (0.1-2 keV) has revealed a rich emission spectrum dominated by narrow emission lines and radiative recombination continuum (RRC) features, never previously observed this clearly in \object{NGC 5548}.

This paper is one of a series exploring different aspects of the 2013-14 observational campaign \citep{Mehdipour:2014ip}. Here we focus on the revealed soft X-ray narrow emission lines and RRC features, as they are excellent probes of the physical characteristics of the emitting region.

\subsection{AGN emission features}
 
There are two main emission line regions thought to occur around AGN systems; the broad-line region (BLR) and narrow-line region (NLR). 
In \object{NGC 5548}, the H$\beta$-emitting BLR extends from at least $\sim4$ to $\sim27$ light days and the higher-ionization lines arise even closer to the central source \citep[e.g.,][]{2004ApJ...613..682P,Bentz:2009kb}.
While the NLR is sufficiently large to be spatially resolved in some AGNs, it is unresolved in \object{NGC 5548}. However, \cite{Peterson:2013iz} measure a size for the [O III]-emitting region of $1-3$ pc based on emission-line variability, consistent with the photoionization prediction of \cite{Kraemer:1998he}. Similarly, \cite{Detmers:2009bq} estimate a radius of $1-15$ pc for the X-ray NLR.

In some cases the high ionisation emission lines from AGN (of which the X-ray narrow lines would be part) are referred to as originating from the ``coronal line region'', distinct from the ``standard'' optical NLR (characterised by [O III]) by the higher ionisation species and larger line widths \citep[see e.g.,][]{1997A&A...323..707E,2002ApJ...579..214R}.
On the other hand, the spatial extent and morphology of the optical and the X-ray NLRs have been found to be remarkably similar in some sources \citep{Bianchi:2006js}.
In this work we continue to distinguish between the X-ray and optical NLRs, instead of using the term ``coronal line region'' to refer to the narrow X-ray emission lines.

Narrow emission lines from AGN are historically thought to remain constant in flux, but there is growing evidence that, in at least some cases, these lines vary over timescales of years. Variability has been presented by \cite{Detmers:2009bq}, using O VII f (forbidden line) in the X-rays, and both \cite{Peterson:2013iz} and \cite{Denney:2014uk}, using [O III] in the optical (the papers refer to \object{NGC 5548} in the former two cases and \object{Mrk 590} in the latter).

RRC features are from free electrons directly recombining into the bound state of an ion. The free electrons involved in this process can have a range of energies, as long as they are above the transition threshold, therefore the RRC features are asymmetrical and can be very broad. The width of these features gives a direct measurement of the energy of the electrons and therefore of the temperature of the plasma. Broad (narrow) RRCs indicate a large (small) range of electron energies above the recombination threshold, and therefore imply high (low) temperatures.
RRC widths have previously been used to determine the temperature of the emitting gas in both obscured Seyfert 1 \citep[e.g.][]{Nucita:2010fn} and traditional Seyfert 2 spectra \citep[e.g.][]{Kinkhabwala:2002fx}.

Narrow emission triplet lines from He-like ions are used as diagnostic tracers of the physical state of the emitting region. In particular, the $G = (f + i)/r$ and $R = f/i$ ratios \citep{Porquet:2000cv} between fluxes of the forbidden ($f$), intercombination ($i$) and resonance ($r$) lines have been applied to give information on the temperature and density of emitting plasmas, although concerns on the limitations to their use in photoionised plasmas have been expressed by \cite{2007ApJ...664..586P} and \cite{Mehdipour:2015ex}.

Here we use the ionisation parameter $\xi \equiv L/nr^{2} $ where $L$ is the ionising luminosity in the 1-1000 Ryd band, $n$ is the hydrogen number density and $r$ the distance from the ionising source \citep{Tarter:1969km}. The parameter $\xi$ is used to characterise components of the warm absorbers (WAs), which imprint absorption features in the soft X-ray band. 

Several authors have reported results using the photoionisation code {\sc Cloudy} to model narrow emission line data from \xmm RGS observations of both Seyfert 2 and low flux states of Seyfert 1 galaxies.
A range of results are summarised in Table \ref{otherauthors_table}.
The authors quoted used a different formulation of ionisation parameter to that adopted in the present work. They used $U \equiv Q/4\pi r^2cn_H$ where $n_H =$ hydrogen number density, $r =$ radius to the inner face of the gas, $Q =$ number of hydrogen-ionising photons emitted per second by a central source of luminosity $L$ and $c$ is the speed of light. $U$ and $\xi$ cannot be directly compared without taking the spectral energy distribution (SED) of the central source into account, therefore here we have quoted the values of $U$ from each work.

\begin{table*}
\begin{minipage}[t]{\hsize}
\setlength{\extrarowheight}{3pt}
\caption{Previously reported results of Seyfert galaxy narrow line region modelling using {\sc Cloudy}}
\label{otherauthors_table}
\centering
\renewcommand{\footnoterule}{}
\begin{tabular}{c c c c | c c c}
\hline \hline
& & & & \multicolumn{3}{c}{{\sc Cloudy} parameters of emitting gas} \\
Reference & Galaxy & Type & Wavelength Band & No. of & Ionization & Density (cm$^{-3}$) \\
& & & & Phases & $\log$\,U & $\log$\,$n_{H}$ \\ 
\hline
\citealp{Kraemer:1998he} & \object{NGC 5548} & Seyfert 1 & UV & 2 & -1.5 & 7 \\
& & & & & -2.5 & 4 \\
\hline
\hline
& & & & \multicolumn{3}{c}{{\sc Cloudy} parameters of emitting gas} \\
Reference & Galaxy & Type & Wavelength Band & No. of & Ionization & Column Density (cm$^{-2}$) \\
& & & & Phases & $\log$\,U & $\log$\,$N_{H}$ \\ 
\hline
\citealp{Armentrout:2007bh} & \object{NGC 4151} & Seyfert 1.5 & X-ray & 3 & 1.3 & 23 \\
& & & & & 0 & 23 \\
& & & & & -0.5 & 20.5 \\
\hline
\citealp{Longinotti:2008di} & \object{Mrk 335} & Seyfert 1 & X-ray & 2 & 0.4 & $\leq 22$ \\
& & & & & 0.8 & $\leq 22$ \\
\hline
{\citealp{Guainazzi:2009fv} \footnote{These authors also found strong evidence for a contribution from collisional components, related to the known ring of starburst activity in this object and not specified here}} & \object{NGC 1365} & Seyfert 1.8 & X-ray & 1 & 1.6$^{+0.3}_{-0.4}$ & $\geq 22$ \\
\hline
\citealp{Nucita:2010fn} & \object{NGC 4051} & Seyfert 1 & X-ray & 2 & 0.6 $<\,\log\,$U$\,<$ 1.9 & 21.77-22.72 \\
\hline
\citealp{Marinucci:2010ba} & \object{NGC 424} & Seyfert 2 & X-ray & 2 & 0.23 $<\,\log\,$U$\,<$ 1.41 & 21.77-22.11 \\
\hline
\hline
\end{tabular}
\end{minipage}
\end{table*}

In this paper we first detail the observations used (Sect. \ref{obs_sect}) and data analysis techniques (Sect. \ref{analysis_sect}), including the He-like triplet diagnostics (Sect. \ref{triplet_sect}). The overall best fit spectrum is presented in Sect. \ref{bestfit_sect}. We then build a consistent photoionisation model using {\sc Cloudy} (Sect. \ref{cloudy_sect}).  The physical implications of our analysis are discussed in Sect. \ref{discussion_sect}. Finally, the conclusions of this work are summarised in Sect. \ref{conclusions}.

\section{Observations}
\label{obs_sect}

\begin{figure*}[!]
\centering
\resizebox{\hsize}{!}{\includegraphics[angle=90]{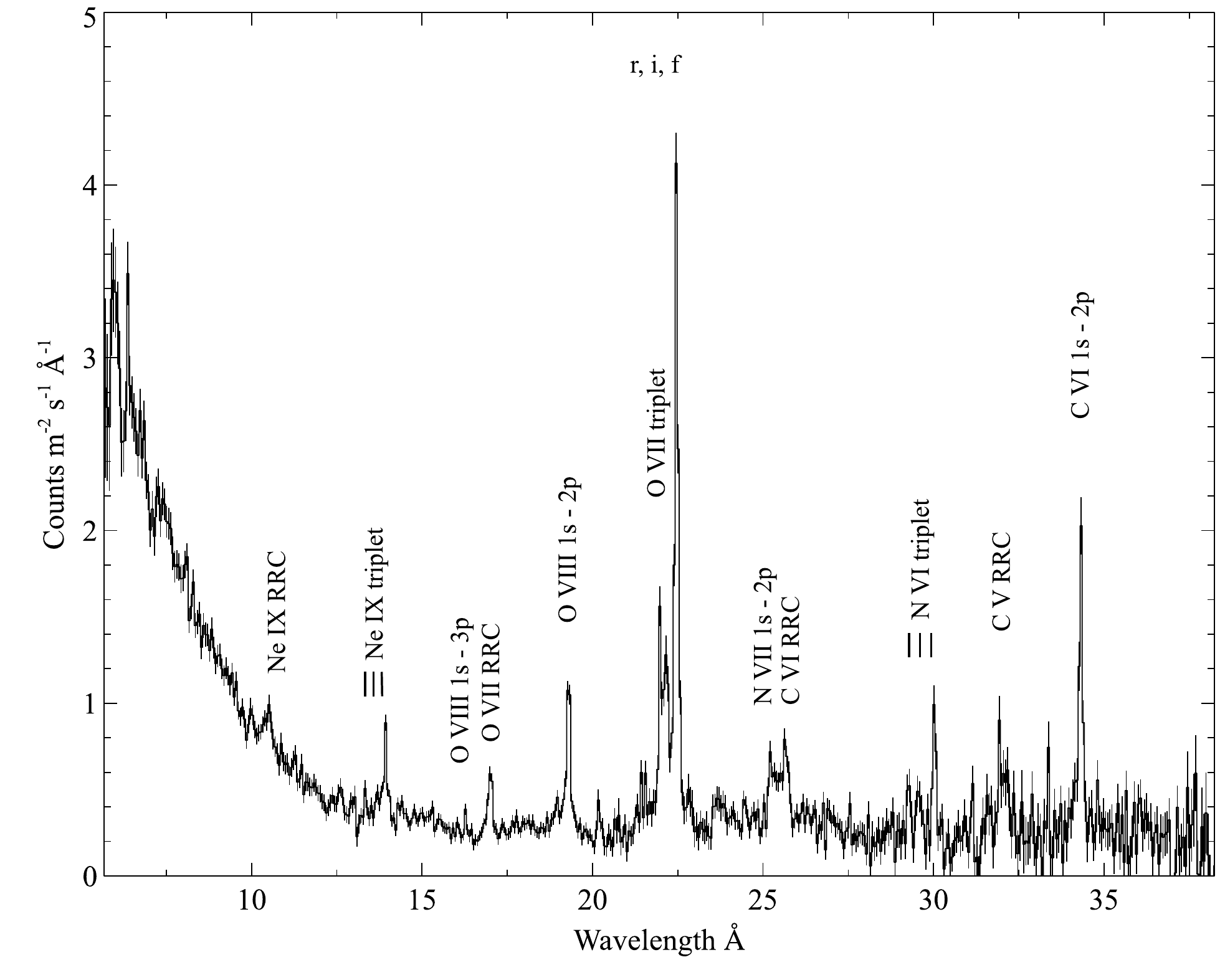}}
\caption{The 2013-14 770\,ks stacked RGS spectrum of \object{NGC 5548} (observed frame). The strongest emission lines are labelled. The data points are shown in black. The data have been binned by a factor of two for clarity.}
\label{all_data}
\end{figure*}

The data used for this paper were taken as part of a large multi-wavelength campaign aimed at characterising in detail the WA in \object{NGC 5548}; the structure, data reduction techniques used and other results of this campaign are explained in \cite{Kaastra:2014ip} and \cite{Mehdipour:2014ip} (the latter includes a log of all observations taken). All spectra shown in this paper are in the observed frame.

Where necessary we have used the redshift value $z = 0.017175$, determined from the 21\,cm Hydrogen line \citep{1991rc3..book.....D}, as given in the NASA/IPAC Extragalactic Database (NED), which is slightly different to some previous work that used a redshift value $z = 0.01676$ determined from the optical lines \citep[e.g. used in][]{Steenbrugge:2003be}.

Here we present a study of the narrow emission lines in the spectra of \object{NGC 5548} obtained by the RGS instrument onboard \xmm \citep{denHerder:2001fv}. We focus on a 770$\,$ks spectrum in the range 5.7-38.2$\,\AA$ (see Fig. \ref{all_data}), created from combining fourteen observations of $\sim$50$\,$ks; twelve of these were taken each a few days apart between 22 June and 31 July 2013, with two other \xmm observations taken a few months later (December 2013 and February 2014).

The fourteen individual observations have also been examined separately, by fitting each narrow emission line in each spectrum with an individual gaussian, and show no significant variability in flux or width of the narrow emission features. Therefore analysing the stacked spectrum of 770$\,$ks for the remainder of this work is justified in order to increase data counts and signal to noise levels within the spectrum.

The RGS data processing was performed at a more advanced level than the standard SAS pipelines; the details of the data reduction techniques used are reported in \cite{Kaastra:2011bo} and summarised in relation to this campaign in \cite{Mehdipour:2014ip}.

The long wavelength part of the spectrum (in Fig. \ref{all_data}, around 35-38$\,\AA$) has noticeably larger uncertainties, due to the decrease in effective area of RGS in this range \citep{denHerder:2001fv}. There is also a decrease in effective area at the short wavelength part of the spectrum (5-10$\,\AA$) but the much higher photon count from the source at these wavelengths slightly reduces the effect this has on the observed spectrum.

\section{Data analysis}
\label{analysis_sect}

Due to the low soft X-ray continuum flux, narrow emission lines (including He-like triplets of Oxygen, Nitrogen and Neon) and radiative recombination continuum (RRC) features dominate the 2013-14 spectrum of \object{NGC 5548}. These features have not been seen so clearly in this source during previous observations, when the soft X-ray continuum was higher \citep[e.g.][]{Kaastra:2000uq,Steenbrugge:2003be}.

In order to accurately analyse the narrow emission features of this spectrum, a good model for the continuum and absorption was needed. \cite{Kaastra:2014ip} conclude that the classical WA previously observed in \object{NGC 5548} is not sufficient to explain the unexpected degree of absorption shown in this spectrum, and identify two new ``obscurer'' components needed to fit the intrinsic and reflected continuum well, in addition to six WA components (named from A to F; parameters shown in Table \ref{WA_table}). The continuum was modelled with a combination of a modified black body and power-law, including reflection from distant material and an exponential cut-off at high energies. The details of this are in the supplementary materials of \cite{Kaastra:2014ip}. The exact physical interpretation of this continuum does not affect this work on the narrow emission features as it is used only as a continuum level above which we measure the emission features. Galactic neutral absorption is included
with H I column density in the line of sight set to $N_H = 1.45$x$10^{20}\,$cm$^{-2}$ \citep{Wakker:2011ea}. This continuum and absorption model is used throughout the analysis in this paper.

Due to low numbers of counts per bin,
 $\chi^{2}$ is not the appropriate goodness-of-fit parameter. Instead, goodness-of-fit should be measured by minimising the C-statistic \citep{Cash:1979bi} parameter as it is based on the Poisson distribution and so does not suffer from the same drawbacks when using low count statistics. We therefore use the C-statistic for all analysis in this work.
The errors used in this paper are all at the 1$\sigma$ (68\%) confidence level unless stated otherwise.

The high spectral resolution of RGS ($\lambda / \Delta\lambda\sim340$ at $22\,\AA$; \citealp{denHerder:2001fv}) compared to lower resolution X-ray spectroscopy leads to more detailed models being needed to fully describe the data. Our knowledge of many aspects of these models is not yet complete (for example presence of additional astrophysical components in the line of sight towards the source and imperfect atomic data) and we also have to deal with imperfect instrumental calibration, systematic effects and statistical fluctuations, so there will necessarily be differences between the data and the best physical model that can be found. As discussed in \cite{Blustin:2002hj}, the use of goodness-of-fit statistics when modelling RGS data must be coupled with careful consideration of the physical meaning of the models, as physical self-consistency is more important than reaching a nominally ``acceptable'' goodness-of-fit value.

All the spectral fitting in this paper was done with the {\sc spex}\footnote{http://www.sron.nl/spex} spectral fitting program \citep{1996uxsa.conf..411K} version 2.05.02.
Additional analysis compares simulated photoionisation spectral models from {\sc Cloudy} \citep[version 13.03;][]{2013RMxAA..49..137F} with the observational data (Sections \ref{cloudy_sect} and \ref{Cloud_geometry_testing}).
The adopted cosmological parameters in our modelling are $H_0 = 70$ \kms Mpc$^{-1}$, $\Omega_{\Lambda}=0.70$ and $\Omega_{m}=0.30$.

%
\subsection{Broad emission lines}
\label{BL_sect}

\begin{table*}
\begin{minipage}[t]{\hsize}
\setlength{\extrarowheight}{3pt}
\caption{Broad emission line parameters from the 2013-14 RGS spectra. All values without error bounds were fixed during fitting. The outflow velocities and widths of both O VII lines were chosen to be the same as those used by the \cite{Steenbrugge:2005dr} analysis of 2002 LETGS data.}
\label{BL_table}
\centering
\renewcommand{\footnoterule}{}
\begin{tabular}{l c | c c c c }
\hline \hline
Line & Rest $\lambda$ & {$\lambda$ \footnote{measured values in the rest frame}} & Outflow Velocity & FWHM & Luminosity \\ 
& ($\AA$) & ($\AA$) & (\kms) & ($\AA$) & (10$^{33}\,$W) \\ 
\hline
O VII 1s-3p & 18.63 & 18.63 & 0 & 0.50 & $<$1.8 \\
O VIII 1s-2p & 18.97 & 18.89$\pm$0.06 & $-1200\pm$900 & 0.12$\pm$0.08 & 10.3$_{-7.4}^{+23.9}$ \\
O VII 1s-2p & 21.60 & 21.60 & 0 & 0.58 & 8.5$\pm$4.4 \\
C VI 1s-2p & 33.74 & 33.65$\pm$0.03 & $-800\pm$300 & 0.24$^{+0.12}_{-0.05}$ & 32.6$\pm12.1$ \\
\hline
\hline
\end{tabular}
\end{minipage}
\end{table*}

\begin{figure}[!]
\centering
\resizebox{1.0\hsize}{!}{\includegraphics[angle=0]{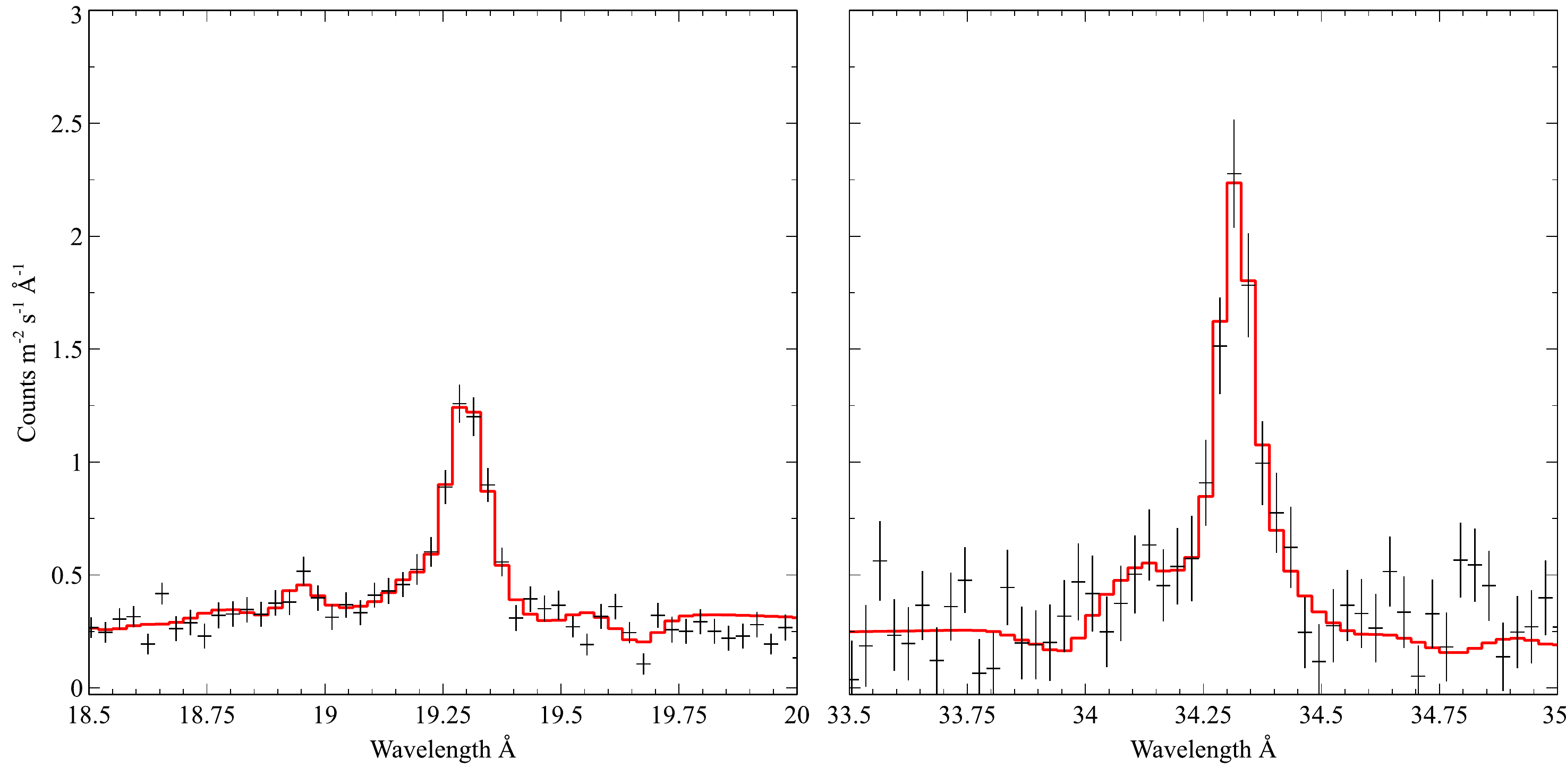}}
\caption{The two broad emission lines that are well fitted; O VIII Ly-$\alpha$ (18.5-20\,$\AA$) and C VI Ly-$\alpha$ (33.5-35\,$\AA$). Both can be seen on the blue side of their corresponding narrow emission line. See text for details.}
\label{broad_line_fits}
\end{figure}

Although broad emission lines are difficult to identify in the 2013-14 spectrum, \cite{Kaastra:2014ip} include four of these in their model (C VI Ly-$\alpha$, O VII He-$\alpha$, O VII He-$\beta$, O VIII Ly-$\alpha$). All of their parameters (FWHM, luminosity and outflow velocity) were initially taken from the \cite{Steenbrugge:2005dr} analysis of the \chandra LETGS 2002 spectrum, when the continuum was higher, and we begin by setting the broad lines at those values, using a \gaussian model for each. The 2002 parameters were estimated from knowledge of the UV broad line parameters in this source as they could not be directly fitted.
We then allowed these parameters (FWHM, outflow velocity and luminosity of each line) to vary so we could fit them to the 2013-14 770\,ks spectrum where possible (see Table \ref{BL_table}).
We take the broad line source to be interior to the obscurer, as expected from the results of \cite{Kaastra:2014ip}, therefore (as the fitted model parameters are different to the 2002 analysis) the new luminosities differ from those found previously. We can fit the broad O VIII Ly-$\alpha$ line (over the range 18.5-20$\,\AA$; C-stat $= 80$, and d.o.f. $= 41$) and the broad C VI Ly-$\alpha$ line (33.5-35$\,\AA$; C-stat $= 75$, and d.o.f. $= 44$), both shown in Fig. \ref{broad_line_fits}.

For both of the O VII broad lines (He-$\alpha$ and He-$\beta$), we assume their FWHM and outflow velocity values have not changed since the values derived for the 2002 spectrum due to difficulties fitting new values in the 2013-14 spectrum. Therefore these two lines are set to their rest wavelength (i.e. assuming no outflow velocity), as they are in \cite{Steenbrugge:2005dr}. For the O VII He-$\alpha$ line this gives us a luminosity estimate, but for the O VII He-$\beta$ this only gives an upper limit to the luminosity (see Table \ref{BL_table}).
As we only have an upper limit, we do not include the broad O VII He-$\beta$ line in any further analysis.

The parameters for the O VIII Ly-$\alpha$, O VII He-$\alpha$ and C VI Ly-$\alpha$ lines are fixed at the values in Table \ref{BL_table} throughout all the rest of the analysis in this paper.

%
\subsection{Initial fit and He-like triplet diagnostics}
\label{initialfit_sect}

\begin{table*}
\begin{minipage}[t]{\hsize}
\setlength{\extrarowheight}{3pt}
\caption{Initial parameters for the O VII triplet from the 770\,ks spectrum. C-statistic (21.5-23\,$\AA$) $=$ 48 for d.o.f. $=$ 49}
\label{initialfit_table}
\centering
\renewcommand{\footnoterule}{}
\begin{tabular}{l | c c c c c }
\hline \hline
{Line \footnote{r, i and f represent the resonance, intercombination and forbidden lines, respectively}} & Rest $\lambda$  & {Measured $\lambda$ \footnote{\object{NGC 5548} rest frame values} \footnote{There is an additional $\pm0.004\,\AA$ error on these values due to RGS absolute wavelength uncertainty}} & Flow velocity & FWHM & FWHM \\ 
& ($\AA$) & ($\AA$) & (\kms) & ($\AA$) & (\kms) \\ 
\hline
O VIIr & 21.602 & 		21.601$\pm$0.004 & $-20\pm50$ & 						$<$0.002 & $<$30 \\
O VIIi & 21.807 & 		21.772$\pm$0.020 & $-480\pm160$ & 					0.151$\pm$0.024 & 2100$\pm$330 \\
O VIIf & 22.101 & 		22.077$\pm$0.002 & $-320\pm40$ & 					0.064$\pm$0.007 & 870$\pm$100 \\
\hline
\hline
\end{tabular}
\end{minipage}
\end{table*}

There are He-like emission line triplets of Oxygen, Nitrogen and Neon within the \object{NGC 5548} stacked spectrum. The O VII triplet is the strongest and the O VII forbidden line component is the highest flux emission line in the data.
For this reason, we have concentrated our diagnostics on the O VII triplet.

All three emission lines within any triplet (four if taking into account the intercombination line as an unresolved doublet) are expected to have consistent velocities. When fitting the He-like Oxygen triplet by using three simple \gaussian models in {\sc spex}, we found the forbidden and intercombination lines to have outflow velocities of $-320\pm40\,$\kms and $-480\pm160\,$\kms respectively, while the resonance line had a best fit velocity consistent with being at rest ($-20\pm50\,$\kms; all values shown in Table \ref{initialfit_table}).

First we looked at whether calibration issues could have caused this apparent difference by comparing with a 68\,ks exposure of Capella taken close to our data, on 31-08-2013 (ObsID 0510781201), and processed the same way as our \object{NGC 5548} data. We compared the line centroids of 8 strong emission lines between 6.64 and 33.74$\,\AA$ with their laboratory wavelengths and found the weighted mean difference to be $-2.7\pm0.4\,$m$\AA$, 
corresponding to $-17\pm2\,$\kms at 20$\,\AA$. This very small offset is compatible with the typical orbital velocity of 30 \kms of the binary components of Capella, and shows that the uncertainties on the absolute wavelength scale are much smaller than the offsets we find for the forbidden and intercombination lines in our \object{NGC 5548} data. In the Capella data, between 15 and 25$\,\AA$, the centroids of the five strongest lines were consistent within the error bars with the weighted mean and all had uncertainties of less than 30$\,$\kms.
We conclude that there is no instrumental reason to doubt the measured line centroids of the O VII triplet in the \object{NGC 5548} spectrum.

Further evidence that this velocity discrepancy is a real effect can be found in historical data; the same apparent velocity difference is seen in 1999 \chandra data, analysed by \cite{Kaastra:2000uq} and \cite{Kaastra:2002ic}, where the O VII resonance line appears $\sim$300$\,$\kms redshifted with respect to the O VII forbidden and intercombination lines, i.e. displaced in the same direction as we find in the 2013-14 spectrum. This previous work focussed on the narrow absorption lines that could be seen in the data, therefore those authors did not investigate the velocity discrepancy between these emission lines.

\subsection{Investigation into O VII triplet velocity discrepancy}
\label{triplet_sect}

After ruling out calibration effects, we investigated various scenarios to explain the velocity discrepancy between the O VII triplet lines. 
First we explored the case in which we may be observing the combined emission from two ionised cones propagating out from the central source in opposite directions. Then we investigated the possibility that the velocity discrepancy is due to absorption.

\subsubsection{Two cone model}
\label{2cone_sect}

First we looked at the possibility that the velocity discrepancy between the O VII triplet lines is caused by the presence of two contributing O VII triplets of different line ratios, one redshifted and one blueshifted with respect to the host galaxy. This could be the case if the narrow emission lines are being produced in two opposing cone shaped regions, each illuminated by and expanding away from the central source, with more gas in the cone closer to (and outflowing towards) the observer. In this case, the optically thin lines (such as forbidden and intercombination lines) would be seen from all of the gas and therefore would have one average (measured) velocity, dominated by that of the closer region with more gas. If the optically thick lines (such as the resonance line) are mainly produced at the illuminated face of the gas and shielded by the cooler gas, then we would preferentially see the lines formed in the gas which faces towards the observer (the further cone from us, outflowing away from us, and therefore redshifted). This would cause an apparent velocity difference between the resonance and both the intercombination and forbidden lines.

We tested this theory in {\sc spex} by modelling each narrow emission line as a combination of two separate \gaussian components, one blueshifted and one redshifted, to represent the two opposing conical regions.
The velocity difference between these components must be small enough that the observed emission lines retain their single peak appearance; this restriction is more important at longer wavelengths. The narrow emission line we observe in our data with the longest rest wavelength is C VI 1s-2p (33.736$\,\AA$), therefore the single peak appearance of this line in our data sets a limit on the velocity difference between the two conical regions in this model.

The best fit to the C VI 1s-2p line had ouflow velocities of $-305\pm40$ and $+125\pm50$\,\kms for the blueshifted and redshifted components respectively. This model (with the outflow velocities fixed to those values) was then extended to fit all the emission lines in the spectrum, one by one, leaving widths and normalisations free.
Finally the outflow velocities of the \gaussian components were left free to vary
to find the final best fit for this model. The final best fit outflow velocities in this case are $-320\pm30$ and $+190\pm30$\,\kms respectively, with a C-statistic of 1832 (d.o.f. $=$ 1027).

\begin{figure}[!]
\centering
\resizebox{1.0\hsize}{!}{\includegraphics[angle=0]{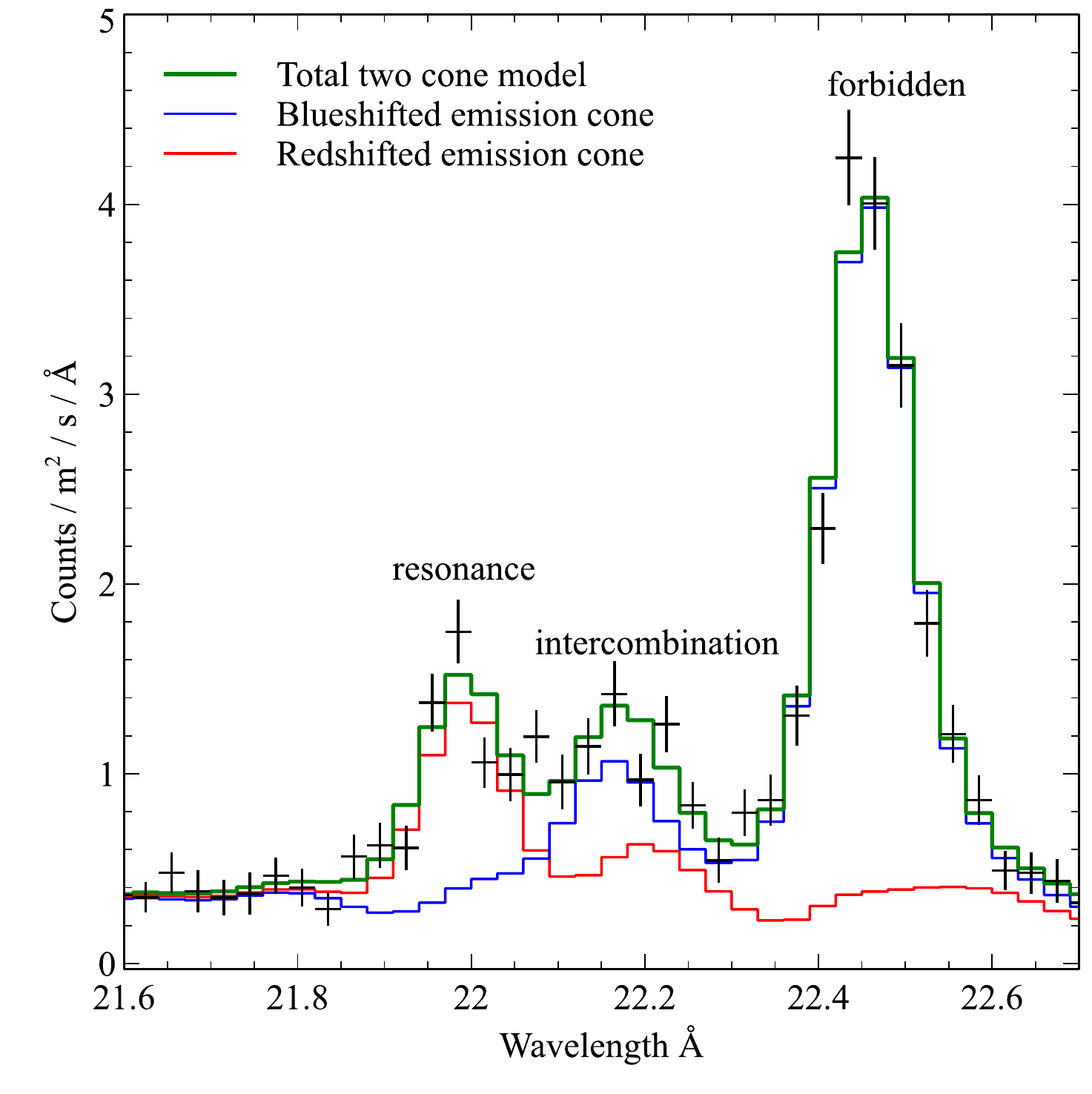}}
\caption{Final two cone model fit of the O VII triplet. Each emission line is fitted with two \gaussian components, one at velocity of $-320\,$\kms (shown in blue) and the other at velocity of $+190\,$\kms (shown in red). The data are shown in black and the overall fit is shown in green. Clearly, the O VII forbidden line (on the right) only requires the blueshifted \gaussian component but this does not match the physical requirements of the two cone model (see text for details).}
\label{OVIInoFline_figure}
\end{figure}

This model is not without problems, as the statistical best fit does not physically match the two cone model explained above. In this model the observed O VII forbidden line should be a combination of O VII forbidden emission from both the blueshifted and redshifted emission cones (as it is an optically thin line), whereas in the best fit model described above, the O VII forbidden line only has a contribution from the blueshifted \gaussian component (see Fig. \ref{OVIInoFline_figure}). In contrast, the Ne IX and N VI forbidden lines in this model do have contributions from both the blue- and redshifted components, as expected. 
As we discuss in Sect. \ref{analysis_sect}, physical self-consistency of our best fit model is the most important factor for acceptance, so for this reason we do not favour the two emission cone model as a physical explanation of the apparent O VII velocity discrepancy.

For completeness, we do look into this scenario further in Sect. \ref{Cloud_geometry_testing} when investigating possible geometries for the narrow-line emitting gas using self-consistent photoionization models from {\sc Cloudy} (this is not tried with {\sc spex} as it does not yet include photoionisation emission models).

\subsubsection{Absorption}
\label{abs_sect}

We then explored the possibility that the shift in the resonance line with respect to the other components of the O VII triplet is due to absorption on its blue side. This could then allow the true velocity of the resonance line to match the outflow velocity of around $-300\,$\kms measured for the intercombination and forbidden lines. The resonance line could be more strongly affected by absorption than the other components of the triplet due to resonant scattering and the possible geometry of the region \citep{2010SSRv..157..103P}.
The comparative widths of these three lines, when simply fitted with Gaussian models, are already an indication that absorption may be playing a role in the appearance of the resonance line. As shown in Table \ref{initialfit_table}, the measured width of the O VII resonance line is extremely small, especially compared to the intercombination and forbidden lines within that same O VII triplet. As all three of these lines should have the same widths, this implies that part of the resonance line is not being observed for some reason; absorption could cause this effect.

\begin{figure}[!]
\centering
\resizebox{\hsize}{!}{\includegraphics[angle=0]{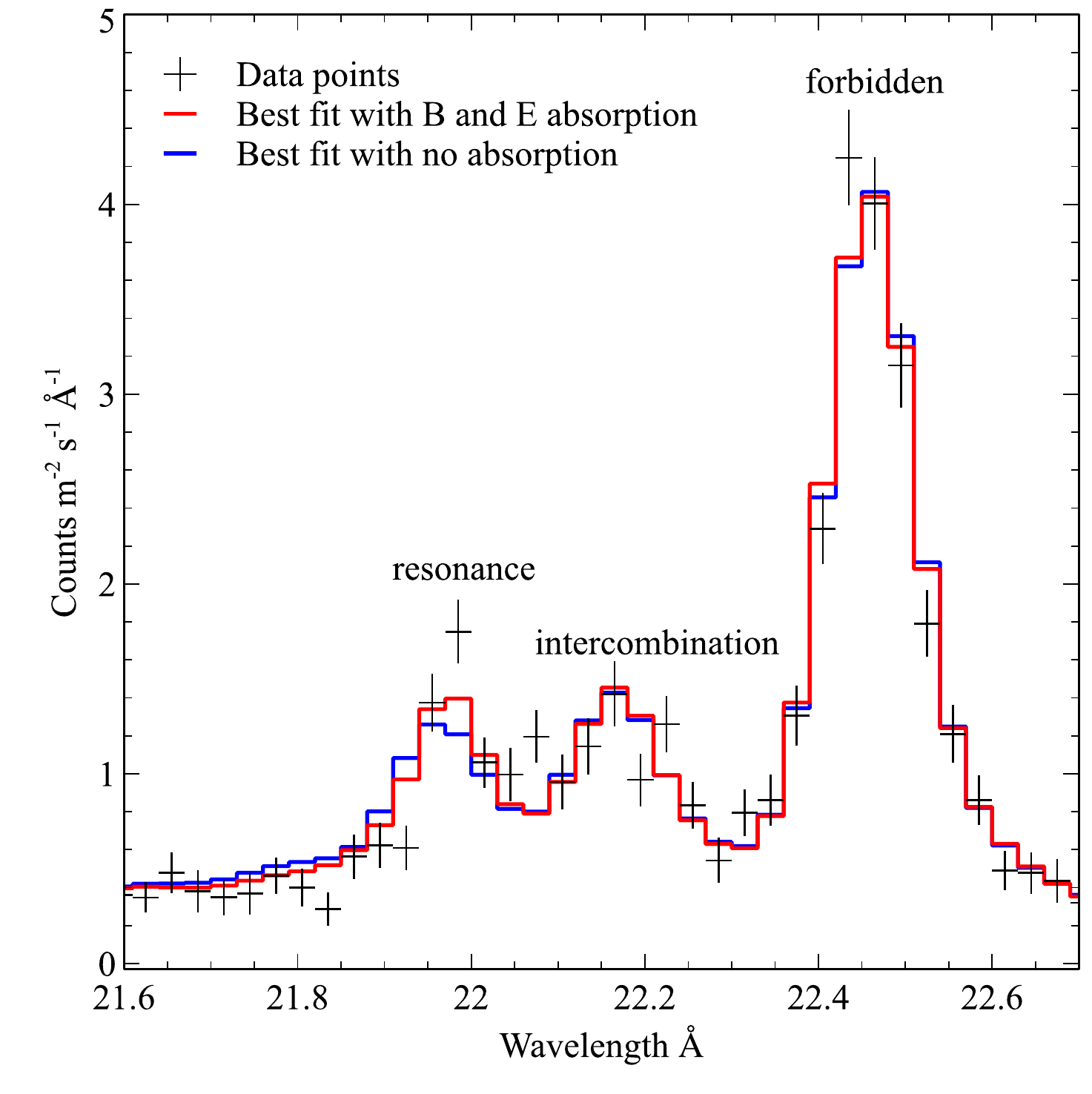}}
\caption{Best fit model to O VII triplet. The data are shown in black. Shown in red is the best fit including absorption from X-ray WA components B \& E (C-statistic $=$ 82, for 43 d.o.f). Shown in blue is the best fit with no absorption (C-statistic $=$ 104, for 43 d.o.f). In both models the outflow velocities and line widths of all three triplet lines are coupled together. The outflow velocity, line width and normalisations were left free to vary to find the best fit for each scenario.}
\label{OVII_68}
\end{figure}

We tested whether absorption of the triplet by any of the six WA components applied by \cite{Kaastra:2014ip} could explain the shift of the resonance line in this way, as all WA components are found to have blueshifted velocities. Previous distance estimates for both the X-ray WA components (at least one component $<$7\,pc; \citealp{Detmers:2008hf}) and X-ray NLR (1-15\,pc; \citealp{Detmers:2009bq}) are similar, but the ``obscurer'' components discovered by \cite{Kaastra:2014ip} are known to be located much closer to the ionising source than the NLR. Therefore these ``obscurer'' components were not included in this absorption test. Parameters of the WA components can be seen in Table \ref{WA_table}. For this test the normalisations of the three O VII triplet lines were left free to vary after being placed inside each of the WA components in turn. The FWHM values for each of the O VII triplet lines were coupled and left free to vary. The outflow velocities for the three O VII triplet lines were also coupled and initially fixed to $-300\,$\kms (the approximate outflow velocity of the forbidden line), to reduce the number of free parameters at this stage. This led to the C-statistic listed in Table \ref{TripAbs_770ks_table_fixedwidth}. The outflow velocity is left free to vary after this test. 

When comparing different, non-nested best fit models with the same number of degrees of freedom, as in this case, standard tests (such as the F-test) are not applicable, so we must use another method to determine if any improvement in the C-statistic is significant. In this case, given the tight relation between $\chi^2$ and C-statistic, we test the significance by taking the square root of the $\Delta$\,C-statistic (measuring the improvement in C-statistic when comparing one model to another) and this gives us the significance of that improvement in $\sigma$. For example, $\Delta$C of $9$ corresponds to a 3$\sigma$ improvement.

\begin{table*}
\begin{minipage}[t]{\hsize}
\setlength{\extrarowheight}{3pt}
\caption{C-statistic values for fitting the O VII triplet, and the whole RGS spectrum, with WA absorption of narrow emission lines (and RRCs when looking at the whole data range). The parameters of all the WA components were fixed to the best-fit values from \cite{Kaastra:2014ip}, as they are for all the analysis within this work. For each of these fits, all three O VII triplet lines had fixed outflow velocities of $-300\,$\kms and the widths of all three triplet lines were coupled and fitted together.}
\label{TripAbs_770ks_table_fixedwidth}
\centering
\renewcommand{\footnoterule}{}
\begin{tabular}{l c c}
\hline \hline
 & C-stat value for & C-stat value for \\  
{Absorption applied \footnote{Letters refer to \cite{Kaastra:2014ip} WA components}} & {O VII triplet \footnote{Data range 21.5-23$\,\AA$, d.o.f. $=$ 43}} & {all RGS data \footnote{Data range 5.68-38.23$\,\AA$, d.o.f. $=$ 1046}} \\
\hline
None & 104 & 1865 \\
A & 103 & 1864 \\
B & 93 & 1856 \\
C & 102 & 1861 \\
D & 114 & 1882 \\
E & 94 & 1879 \\
F & 100 & 1863 \\
\hline
B \& E & 83 & 1870 \\
B, E \& D & 85 & 1878 \\
B, E \& F & 78 & 1871 \\
A, B \& C & 90 & 1853 \\
A, B, C \& D & 96 & 1864 \\
A, B, C, D \& E & 80 & 1866\\
A, B, C, D, E \& F & 78 & 1868 \\
\hline
\hline
\end{tabular}
\end{minipage}
\end{table*}

We focus on the results using only the O VII triplet wavelength range, as the C-stat of this region is not affected by any other regions where the continuum model does not match the data exactly (for example the excess of data over model in the range 35-38\,$\AA$). We conclude that one model is an improvement over another when the C-stat improves by $\geq$9, relating to a $3\sigma$ improvement. 

As shown in Table \ref{TripAbs_770ks_table_fixedwidth} we also calculated the C-statistic values for each scenario over the whole spectrum.
Initially we considered absorption by each WA component in turn; we determined that absorption by WA component B improved the fit substantially, and more so than the other components (from a C-stat of 104 to 93, d.o.f.$=$44). We then tested whether this fit further improved by adding another absorption component and found that including absorption by component E decreased the C-stat (by 10, to 83). Adding a third absorbing component (F) also decreased the C-stat around the triplet range again, but this time only by 5, so we do not consider this enough improvement to justify use of this model over the previous one.
We also tested a different absorption scenario by using components A, B and C to absorb the narrow emission lines. This gave a better C-statistic when fit to the whole spectrum, but a worse value for only the O VII triplet range (see Table \ref{TripAbs_770ks_table_fixedwidth}). As this test is focussed on how well we can fit the O VII triplet, we marginally prefer the scenario using absorption by B and E only because it requires a smaller number of absorbing components and gives a reasonable fit for both the O VII triplet and the whole spectrum. 

We then allowed the outflow velocity of the triplet lines (still coupled together) to vary freely, using absorption from WA components B and E, giving a best fit outflow velocity of $-300\pm30\,$\kms (the C-statistic reduced by 1, to 82). This fit, compared to the fit with coupled O VII triplet line velocities but no absorption, is shown in Fig. \ref{OVII_68}. Clearly, the inclusion of absorption shifts the fitted peak of the O VII resonance line towards the peak seen in the data.

The evidence from these tests so far implies that WA components B and E are doing most of the absorbing of the \mbox{O VII r} line; we decided to explore how important the outflow velocities of these components are to this determination.
To do this we took each of the other WA components and fixed their velocity to that of B and E in turn. Absorption by component A had very little effect on the fit around the O VII triplet, whether it was at its measured velocity or that of component B or E. We put this down to the fact that the measured velocities of components A, B and E are similar ($-588\pm34\,$\kms, $-547\pm31\,$\kms and $-792\pm25\,$\kms respectively, taken from \citealp{Kaastra:2014ip}).
When the other components (C, D and F) were set to the velocities of components B and E, absorption by them caused a substantial improvement in the fit of the O VII triplet, compared to when they were set to their measured velocities. The C-statistic improved by between 5 and 29 (d.o.f. $=$ 44) for each case.

The above indicates that having well determined outflow velocities for these warm absorber components is vital to calculate the effect that absorption by each one would have on the fit of the O VII triplet. Therefore the details of this absorption scenario for the narrow emission lines should be seen as strongly dependent on the determined warm absorber properties; in the context of this work we are testing the concept that the narrow emission lines could be absorbed by some warm absorber components, not the specific order of the involved warm absorber components.

While this absorption scenario does not give a perfect fit, it is the only physically plausible explanation for the O VII velocity discrepancy so far. Therefore we have used absorption by WA components B and E as our best fit scenario for the remainder of this work.

Using this absorption scenario for the O VII triplet, we recalculated the parameters of all narrow emission lines and RRCs (using the \rrc model in {\sc spex}, see Sect. \ref{RRC_sect}) assuming that all are absorbed by the same combination of WA components. The results of this can be seen in Table \ref{TripAbs_770ks_table_fixedwidth}. Absorbing all the narrow lines and RRCs by WA components B and E led to the best fit for the O VII triplet, but not for the whole spectrum. The best fit for the whole data range (5.7-38.2$\,\AA$) is obtained when the narrow lines and RRCs are all absorbed by WA components A, B and C; while this could be another physically acceptable scenario, the improved fit around the O VII triplet when absorbed by WA components B and E leads to our decision to focus on the latter for this work.

The best fit values for the narrow-line and RRC parameters over the whole spectrum, after absorption from WA components B and E, are presented in Sect. \ref{bestfit_sect} and in Tables \ref{NL_table} and \ref{RRC_table} respectively. These parameters are of course dependent on which scenario is used; here we adopt the simplest absorption scenario that gives the best fit to the O VII triplet region (with $\Delta C \geq 9$ from any simpler scenario).

\subsection{{\sc spex} best fit for the 770\,ks stacked spectrum}
\label{bestfit_sect}

\subsubsection{Narrow emission line parameters}
\label{narrow_sect}

\begin{table*}
\begin{minipage}[t]{\hsize}
\setlength{\extrarowheight}{3pt}
\caption{Narrow emission line best fit parameters from the 770\,ks spectrum when absorption from WA components B and E is applied. Parameters without errors were fixed during fitting. Overall C-statistic $=$ 1870 for d.o.f. $=$ 1036.}
\label{NL_table}
\centering
\renewcommand{\footnoterule}{}
\begin{tabular}{l | c c c c c c | c}
\hline \hline
{Line \footnote{r, i and f represent the resonance, intercombination and forbidden lines, respectively}} & Rest $\lambda$  & {Measured $\lambda$ \footnote{\object{NGC 5548} rest frame values} \footnote{There is also $0.004\,\AA$ uncertainty on these values due to RGS absolute wavelength uncertainty}} & {Flow velocity \footnote{The O VII triplet outflow velocity values were coupled together}} & {FWHM \footnote{The Ne IX, O VII and N VI triplets each had the FWHM values of their three lines coupled together}} & FWHM & Unabsorbed Flux & Previous \\ 
& ($\AA$) & ($\AA$) & (\kms) & ($\AA$) & (\kms) & (photons m$^{-2}$ s$^{-1}$) & {detections \footnote{K02 represents \cite{Kaastra:2002ic}. S05 represents \cite{Steenbrugge:2005dr}. D09 represents \cite{Detmers:2009bq}}} \\ 
\hline
Ne X 1s-2p & 12.138 & 	12.138 & - & 										10$^{-4}$ & - & 										$<$0.013 & - \\
Ne IXr & 13.447 & 		13.447 & - &							 				0.056$\pm0.002$ & 1230$\pm$40 & 						0.035$\pm$0.008 & - \\
Ne IXi & 13.553 & 		13.553 & - &										 	0.056$\pm0.002$ & 1230$\pm$40 & 						0.016$\pm$0.007 & S05 \\
Ne IXf & 13.699 & 		13.695$\pm$0.004 & -90$\pm$90 & 						0.056$\pm0.002$ & 1230$\pm$40 &						0.115$\pm$0.008 & K02, S05 \\
O VIII 1s-3p & 16.006 & 	16.003$\pm$0.006 & 0$\pm$110 & 						$<$0.068 & $<$1280 & 									0.049$\pm$0.006 & - \\
O VII 1s-3p & 18.627 & 	18.612$\pm$0.015 & -240$\pm$240 & 					$<$0.086 & $<$1390 & 									0.047$\pm$0.014 & - \\
O VIII 1s-2p & 18.969 & 	18.969$\pm$0.001 & 0$\pm$20 & 						0.079$\pm0.028$ & 1250$\pm$440 & 					0.238$\pm$0.100 & K02 \\
O VIIr & 21.602 & 		21.580$\pm$0.002 & -300$\pm30$ & 					0.082$\pm$0.006 & 1130$\pm$80 & 						0.307$\pm$0.027 & K02 \\
O VIIi & 21.807 & 		21.785$\pm$0.002 & -300$\pm30$ & 					0.082$\pm$0.006 & 1130$\pm$80 & 						0.222$\pm$0.024 & K02, S05 \\
O VIIf & 22.101 & 		22.079$\pm$0.002 & -300$\pm30$ & 					0.082$\pm$0.006 & 1130$\pm$80 & 						0.822$\pm$0.030 & K02, S05, D09 \\
N VII 1s-2p & 24.781 & 	24.806$\pm$0.011 & 300$\pm$130 & 					0.177$\pm$0.028 & 2140$\pm$340 & 					0.169$\pm$0.023 & - \\
N VIr & 28.787 & 		28.753$\pm$0.011 & -350$\pm$110 & 					0.058$\pm$0.014 & 600$\pm$140 & 						0.136$\pm$0.027 & - \\
N VIi & 29.084 & 		29.060$\pm$0.013 & -250$\pm$130 & 					0.058$\pm$0.014 & 600$\pm$140 &					 	0.086$\pm$0.017 & S05 \\
N VIf & 29.534 & 		29.511$\pm$0.006 & -230$\pm$60 & 					0.058$\pm$0.014 & 600$\pm$140 & 						0.176$\pm$0.018 & S05 \\
C VI 1s-2p & 33.736 & 	33.734$\pm$0.005 & -20$\pm$50 & 						$<$0.063 & $<$800 &										0.339$\pm$0.069 & - \\
\hline
\hline
\end{tabular}
\end{minipage}
\end{table*}

There are 14 narrow emission lines detected and an upper limit to one further narrow emission line in the stacked spectrum; their best-fit parameters are in Table \ref{NL_table}. These were obtained using the absorption scenario (B and E absorption) described in detail in Sect. \ref{abs_sect}.
We have checked all the best-fit values we would measure for parameters in Tables \ref{NL_table} and \ref{RRC_table} if we had used absorption from WAs A-C (instead of B and E) and all are within $3\sigma$ uncertanties from the ones we provide. This shows that, for the X-ray narrow lines in this source, choosing a different specific absorption scenario does not significantly alter the overall results. The presence of this absorption is more important than which specific components are causing it.

Three complete emission triplets, of forbidden, intercombination and resonance lines (O VII, N VI \& Ne IX) are present, plus four Ly-$\alpha$ (Ne X, N VII, O VIII \& C VI), O VII He-$\alpha$ and O VIII Ly-$\beta$ lines.
These lines, with a few exceptions, were fitted using the simple \gaussian model in {\sc spex} with the outflow velocity, FWHM and normalisation kept as free parameters.
The exceptions to this are the Ne IX, O VII and N VI triplet lines, which had some values coupled together (see Table \ref{NL_table} for details), and the Ne X Ly-$\alpha$ and the Ne IX resonance and intercombination lines, which were fixed at their \object{NGC 5548} rest frame wavelengths.
The rest wavelengths were all taken from the internal {\sc spex} line list.

Table \ref{NL_table} also shows which of these narrow lines have been previously detected in \object{NGC 5548} \citep[][]{Kaastra:2002ic,Steenbrugge:2005dr,Detmers:2009bq}.

%
\subsubsection{Radiative recombination continuum feature parameters}
\label{RRC_sect}

\begin{table*}
\begin{minipage}[t]{\hsize}
\setlength{\extrarowheight}{3pt}
\caption{RRC parameters for the 770$\,$ks spectrum, including absorption from WA components B and E, as discussed in Sect. \ref{abs_sect}. Overall C-statistic $=$ 1870 for d.o.f. $=$ 1036}
\label{RRC_table}
\centering
\renewcommand{\footnoterule}{}
\begin{tabular}{l l c c c}
\hline \hline
RRC name & Recombining ion & {Emission Measure \footnote{\label{recombining} of recombining ion}} & Temperature & Emission Measure\textsuperscript{ \ref{recombining}} \\
& & for overall fit & for pair fit & for pair fit \\
& & ($10^{64}\,$m$^{-3}$) & eV (K) & ($10^{64}\,$m$^{-3}$) \\
\hline
C V & C VI & 1500$^{+200}_{-190}$ 	& $4.1\pm0.7$ & 1110$\pm200$ \\
C VI & C VII & 900$^{+100}_{-90}$ 	& (4.8$\pm0.8 \times 10^4$) & 620$\pm110$\\
\hline
N VI & N VII & 130$\pm70$ 			& {6.6 \footnote{Fixed to Oxygen best fit value, see text for details}} & 130$\pm80$ \\
N VII & N VIII & 60$\pm30$ 			& ($7.7 \times 10^4$) & 70$\pm30$ \\
\hline
O VII & O VIII & 650$\pm40$ 			& 6.6$\pm0.6$ & 700$\pm50$ \\
O VIII & O IX & 130$\pm20$ 			& ($7.7\pm0.7 \times 10^4$) & 140$\pm20$ \\
\hline
Ne IX & Ne X & 60$\pm10$ 			& 9.4$\pm0.3$ & 80$\pm20$ \\
Ne X & Ne XI & 20$\pm10$ 			& (10.9$^{+0.4}_{-0.3} \times 10^4$) & 30$\pm20$ \\
\hline
& Overall & $5.9\pm0.5$ eV \\
& Temperature & ($6.8\pm0.5 \times 10^{4}$ K) \\
\hline
\hline
\end{tabular}
\end{minipage}
\end{table*}

We detect eight RRC features in this spectrum; the final best fit parameters are presented in Table \ref{RRC_table}. All eight RRC features were simultaneously fitted with one \rrc model in {\sc spex}, which uses one temperature and allows different emission measures for each ion. This first \rrc model gives an electron temperature in eV which is shown (along with the emission measures for each ion) in Table \ref{RRC_table} (third column).
The RRC features were then split into element pairs (e.g. O VII and O VIII together) and fitted with one \rrc model for each pair. The Carbon, Oxygen and Neon \rrc models all give temperatures of the same order of magnitude (Table \ref{RRC_table}). The Nitrogen RRC features are very weak in this spectrum, therefore the temperature for the Nitrogen \rrc model was fixed at the best fit value for that of the Oxygen \rrc model as the O VII RRC is the strongest, non-contaminated RRC feature in this spectrum. Nitrogen RRC emission measure values using this temperature are shown in Table \ref{RRC_table}. Fitting the RRC features in this `paired' way marginally improved the overall C-statistic by $\Delta$C $=$ 10 compared to using just one \rrc model (and therefore one temperature) for all features. The best-fit temperatures for the Carbon and Neon RRC pairs are inconsistent with each other at the $3\sigma$ level (the uncertainties in Table \ref{RRC_table} represent $1\sigma$), which gives some evidence for stratification of the emission regions producing these RRC features.

The threshold energies ($E$) of all the RRC features are of the order of several hundred eV, and the electron temperatures ($\Delta E$) are measured to be of the order of a few eV, therefore $\Delta E/E << 1$ which is strong evidence for photoionised conditions dominating within the emitting gas.

\section{Testing the absorption model using {\sc Cloudy}}
\label{cloudy_sect}

\begin{table*}
\begin{minipage}[t]{\hsize}
\setlength{\extrarowheight}{3pt}
\caption{Results from automated fitting of the 770\,ks stacked spectrum with {\sc Cloudy} simulated spectra. For each fit the {\sc Cloudy} spectra are given an outflow velocity of -300\,\kms, in order to match the O VII f line (the strongest in the spectrum). The emboldened rows show the best-fits to both absorption scenarios when using the fine grid of {\sc Cloudy} models.}
\label{cloudy_table}
\centering
\renewcommand{\footnoterule}{}
\begin{tabular}{c c c c | c c c}
\hline \hline
& & & & \multicolumn{3}{c}{Best fit {\sc Cloudy} parameters} \\
SED & Spectrum range & {Absorption \footnote{Letters refer to \cite{Kaastra:2014ip} WA components}} & C-statistic & $\log\,\xi$ & $\log\,v_{turb}$ & $\log\,N_{H}$ \\
& $\AA$ & & (d.o.f) & & \kms & cm$^{-2}$ \\ 
\hline
obscured & 5.7-38.2 & B, E & 2213 (1084) & 2.3 & 2.5 & 22.5 \\
obscured & 5.7-38.2 & A-C & 2178 (1084) & 2.3 & 2.5 & 22.5 \\
unobscured & 5.7-38.2 & B, E & 2298 (1084) & 1.5 & 2.0 & 23 \\
unobscured & 5.7-38.2 & A-C & 2272 (1084) & 1.5 & 2.0 & 22.5 \\
\hline
unobscured & 21.5-23 & B, E & 97 (49) & 1.7 & 2.5 & 23.25 \\
unobscured & 21.5-23 & A-C & 106 (49) & 1.7 & 2.0 & 22.25 \\
\hline
\textbf{unobscured} & \textbf{5.7-38.2} & \textbf{B, E} & \textbf{2273 (1084)} & \textbf{1.45} & \textbf{2.25} & \textbf{22.9} \\
\textbf{unobscured} & \textbf{5.7-38.2} & \textbf{A-C} & \textbf{2243 (1084)} & \textbf{1.45} & \textbf{2.25} & \textbf{22.9} \\
\hline
\hline
\end{tabular}
\end{minipage}
\end{table*}

In order to test the physical self-consistency of our modelling (as {\sc spex} does not include a photoionised emission model) we use the spectral synthesis program {\sc Cloudy} \citep{2013RMxAA..49..137F}. We created a number of simulated spectra for a range of physical parameters and compared them to our \object{NGC 5548} data using the absorption scenario described above. We assume a plane parallel geometry, with the central source illuminating the inner face of the cloud and with a flux density dependent on a chosen ionisation parameter, $\xi$.

\begin{figure}[!]
\centering
\resizebox{\hsize}{!}{\includegraphics[angle=0]{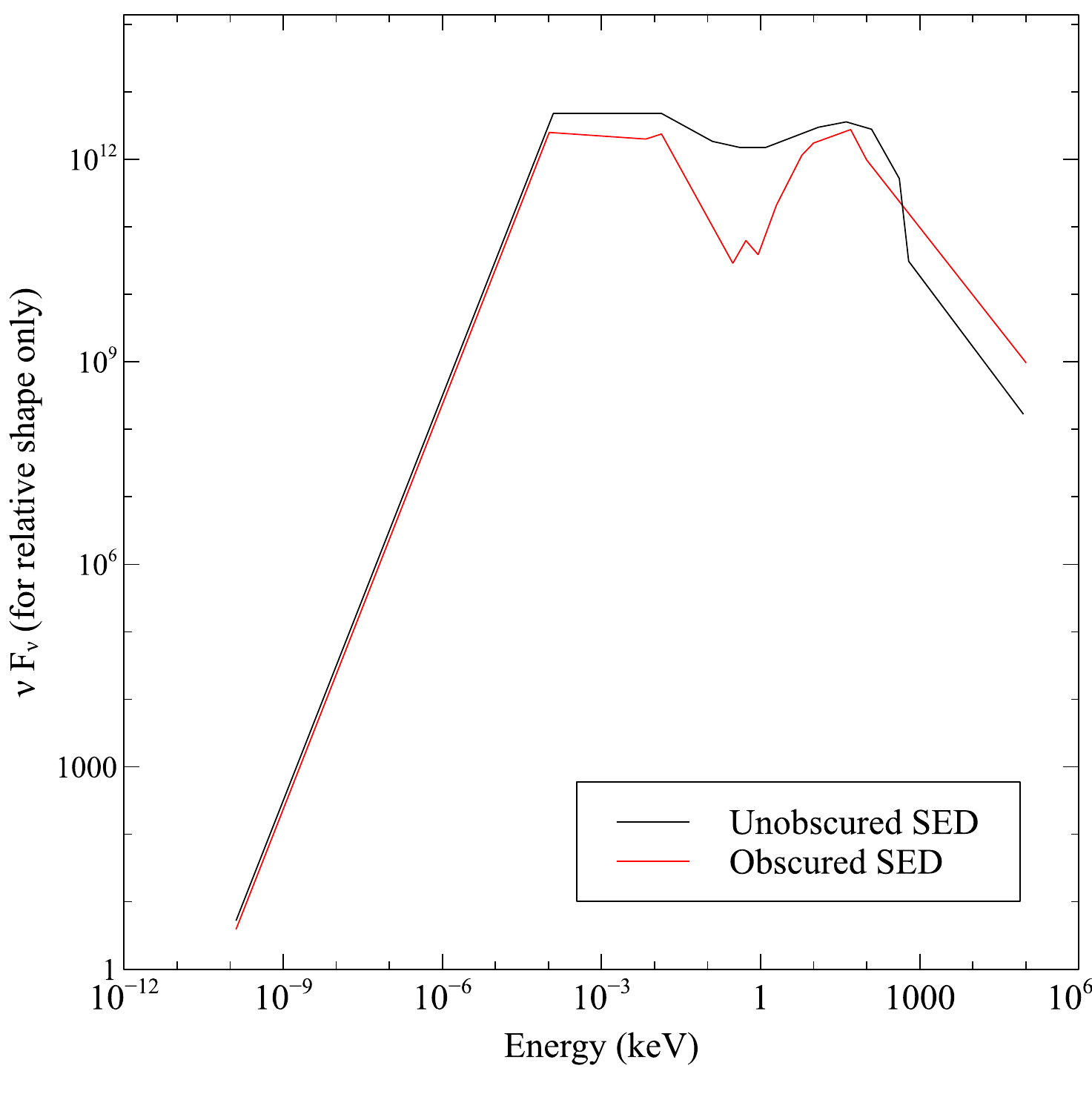}}
\caption{The two SED shapes used as input for {\sc Cloudy}. {\sc Cloudy} scales the input SEDs using either luminosity or intensity values, so absolute values on the y-axis are not relevant.}
\label{cloudy_SED}
\end{figure}

The spectrum produced by photoionisation processes strongly depends on the shape of the ionising continuum, described by the Spectral Energy Distribution (SED). Two sets of {\sc Cloudy} results were compared with the 770\,ks RGS spectrum; one was generated using the historical unobscured SED of NGC 5548 and one used the new obscured SED from 2013-14, both shown in Fig. \ref{cloudy_SED}. These SEDs are smoothed and simplified compared to those presented in \cite{Mehdipour:2014ip}, but qualitatively very similar in shape, which is enough for this analysis. Both sets of {\sc Cloudy} spectra were compared with the data in order to determine which gave the best fit.
If the unobscured SED gives a better fit to the data, this would be evidence that we see the X-ray NLR as still receiving and reacting to the unobscured flux from the central source. If the obscured SED gives a better fit, then this would suggest that the NLR is located close enough to the central source that we see it as already receiving and reacting to the obscured flux.

For both SEDs, 
the {\sc Cloudy} simulated spectra explored a parameter range of log $\xi$ $=$ 0.1-3.3 (with 0.2 step size), hydrogen column density $=$ 10$^{21}$-10$^{24}\,$cm$^{-2}$ (with 10$^{0.25}$ step size) and turbulent velocity $=$ 10$^{1}$-10$^{3.5}\,$\kms (with 10$^{0.5}$ step size). The electron density was set to $10^8\,$cm$^{-3}$ because as long as this remains under the {critical density\footnote{The critical density for any species is the density at which collisional and radiative de-excitation rates are equal; collisional processes become more (less) important for higher (lower) densities.}} for the species of interest ($10^9-10^{10}\,$cm$^{-3}$; see \citealp{Porquet:2000cv} and \citealp{2007ApJ...664..586P} for details) its value is unconstrained by the data we are comparing to.

The 1326 simulated spectra within this parameter grid for each SED were compared with the stacked 770\,ks spectrum.

Each one of the {\sc Cloudy} generated spectra was imported into a {\sc spex} \file model and compared with the data using an automated fitting process, with the normalisation of the simulated spectra left free to vary. This fitting process used a Python script to run {\sc spex} and generate a C-statistic value for each one of the simulated spectra. Throughout this process there is no interpolation between the 1326 {\sc Cloudy} models in the grid specified above.
Unless otherwise stated, the {\sc Cloudy} spectra were all given an outflow velocity of -300$\,$\kms to match the measured value of the strongest emission line in the data, the O VII f line.

There are numerous possible outputs from {\sc Cloudy} because, depending on the geometry of the system, the user may want to model emission in different directions. In terms of emission from the cloud, there are two directional components: reflected and transmitted. The reflected emission is from the illuminated face of the cloud (illuminated by the source), travelling back towards the source. The transmitted emission is the emission from the shielded face of the cloud and travels outwards (away from the source). 

For this first test we chose to use the reflected spectrum output to model the narrow emission lines \citep[as done in e.g.][]{Bianchi:2010dm,Guainazzi:2009fv}.
This is of course only one potential geometry of the NLR region; more geometrical possibilities are explored in Sect. \ref{Cloud_geometry_testing}. 

Each {\sc Cloudy} reflected spectrum includes diffuse emission from within the cloud and scattered (absorbed and re-emitted) radiation from the intrinsic continuum, therefore it includes only the narrow emission lines and a scattered continuum. This means the incident continuum itself, WAs, obscurer and broad emission lines must be modelled separately.
The observed continuum, WAs and obscurer are modelled as in \cite{Kaastra:2014ip} and the 2013-14 broad emission line parameters from Sect. \ref{BL_sect} in this paper are used. We tested whether the continuum should be left free to vary to allow any extra flux from the scattered {\sc Cloudy} continuum to be included in the model without duplicating flux, but this did not change the results. There is not enough scattered continuum in the {\sc Cloudy} spectra for this to be an important factor in the RGS band.

The narrow emission lines {\sc Cloudy} produces are intrinsically narrower than those we see in the RGS data because,
while {\sc Cloudy} takes the effect of turbulence within the gas into account for the line strengths, it does not use this to broaden the lines in the produced {spectra\footnote{See Hazy 1 documentation, page 162}}.
Therefore for the absorption to be applied properly, the narrow emission lines in the {\sc Cloudy} spectra must be broadened. This was done using the \vgau model in {\sc spex}, set to a sigma velocity of 450$\,$\kms as this was manually found to be the best overall fit to the data; it is also consistent with the average Gaussian sigma broadening value ($460\pm30\,$\kms) found for all the narrow emission lines in \cite{Kaastra:2014ip} and overall with the FWHM values in Table \ref{NL_table}.

When fitting the whole 770\,ks spectrum, we found that, with absorption from WA components B and E applied, the {\sc Cloudy} results based on the obscured and unobscured SEDs gave best C-statistic values of 2213 and 2298 respectively (d.o.f $=$ 1084).
All the C-statistic values from the {\sc Cloudy} fits are collected in Table \ref{cloudy_table}.

At this point we compared how well the {\sc Cloudy} models from obscured and unobscured SEDs fit the data. In the RGS data range the {\sc Cloudy} simulated spectra using the obscured SED gave a better fit to the data ($\Delta C = \ \sim$90). We tested to see if this is compatible with the rest of the \object{NGC 5548} spectrum extending to the hard X-rays by applying our best fit model for the RGS band to the EPIC-pn data from the same campaign (see \citealp{Cappi:2015ip} for details of the analysis of the EPIC-pn data). This showed a severe excess of flux in the higher energy ($>$2\,keV) range of the obscured SED {\sc Cloudy} model in comparison to the EPIC-pn spectrum, and leads us to conclude that the {\sc Cloudy} results from the unobscured SED are more physically appropriate, despite the slightly worse statistical quality of the fit in the RGS band.

In order to derive the errors in the best fit parameters for the emitting gas we created a finer grid of models with {\sc Cloudy}, using the unobscured SED. 

From this we calculate uncertainties on the parameters of the emitting gas for both absorption scenarios investigated above (B, E and A, B, C), giving 1\,$\sigma$ uncertainties of log $\xi$ $=$ 1.45$\pm$0.05, log $N_H$ $=$ 22.9$\pm0.4$\,cm$^{-2}$ and log $v_{turb}$ $=$ 2.25$\pm$0.5\,\kms.
It is clear that the ionisation parameter of the emitting gas is much better constrained than the column density (and by extension the turbulent velocity, as this depends strongly on the column density).

\begin{figure*}[!]
\centering
\resizebox{0.95\hsize}{!}{\includegraphics[angle=90]{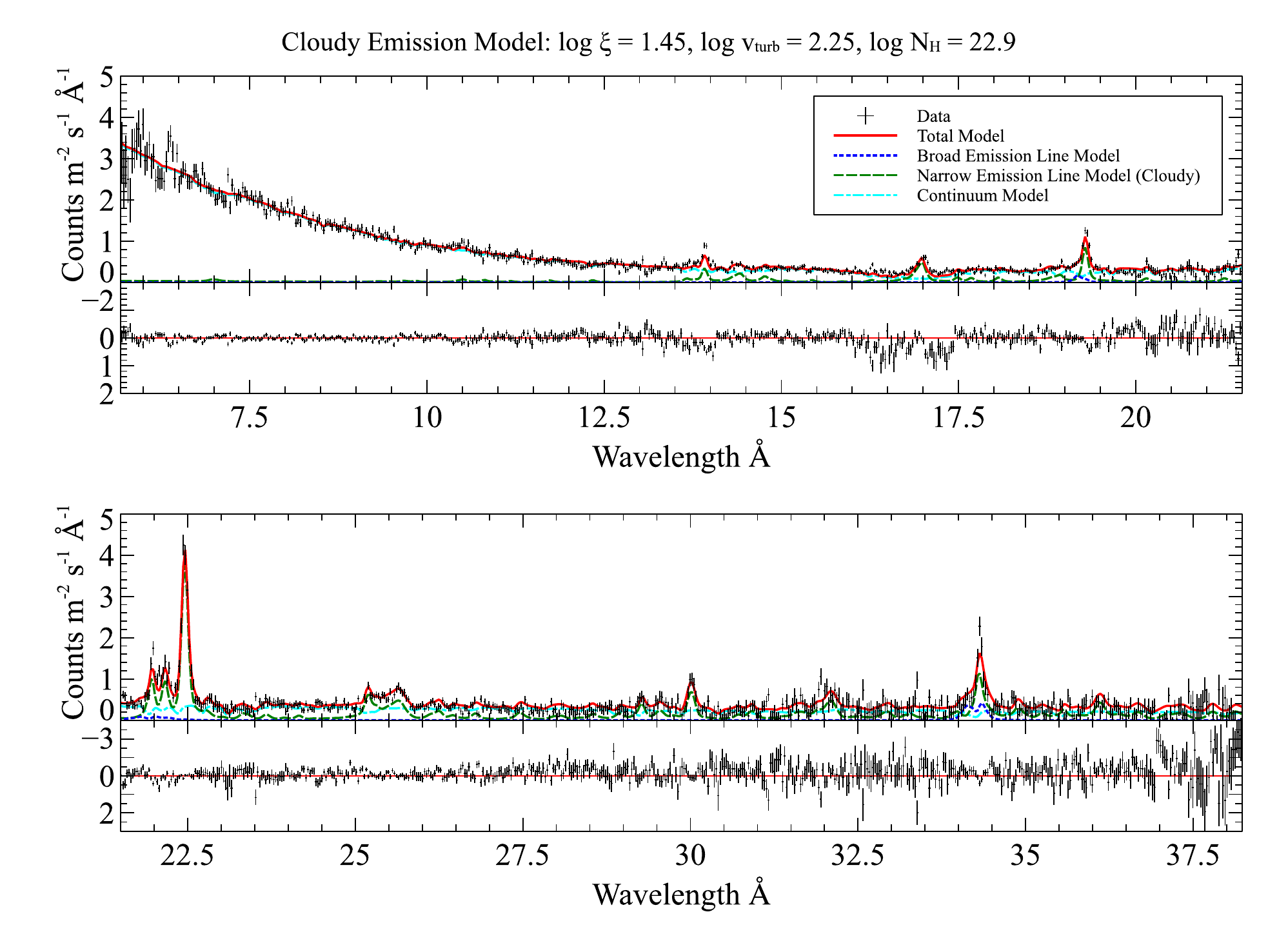}}
\caption{Best fit self-consistent photoionisation model from {\sc Cloudy} for all emission features (5.7 - 38.2$\AA$), including absorption from X-ray WA components B \& E (see \citealp{Kaastra:2014ip} for details). C-statistic: 2273, d.o.f. $=$ 1084. The data are shown in black, the best fit model is shown in red and the components of that model are shown in different colours and line styles.}
\label{all_total}
\end{figure*}

This model is plotted in Fig. \ref{all_total}.

\subsection{Testing further geometries with {\sc Cloudy}}
\label{Cloud_geometry_testing}

In the previous section we have investigated fitting the 770\,ks spectrum with the reflected output from {\sc Cloudy} models. As stated above, this is only one geometrical possibility for the NLR region, and so here we look into other potential geometries using different {\sc Cloudy} {outputs\footnote{For detailed information of {\sc Cloudy}'s geometry see Hazy 1, page 6}}.

In Section \ref{cloudy_sect} we briefly described the outputs that {\sc Cloudy} calculates are produced in different directions from the modelled cloud. The main transmitted (away from the source) output from {\sc Cloudy} contains both the emission from the cloud itself and the attenuated emission from the source; for this work we do not want to include the attenuated emission from the source, so we additively combine the more specific {\sc Cloudy} outputs of transmitted line emission and diffuse emission from the cloud. This approximates the outwards emission from the cloud without including the transmitted attenuated continuum emission, which is not appropriate as we are only using {\sc Cloudy} to model the narrow emission lines, not all components of the total spectrum.

We compare three main different geometries to draw some simple conclusions; this is by no means a full investigation into all possible geometries of this system. These tests were all done with the {\sc Cloudy} grid of models used in the previous section, but with varying outputs used.

Geometry 1)
The simplest interpretation of the reflected output, adopted in the previous section, is that we are observing emission from the illuminated face of NLR clouds on the opposite side of the source to our line of sight.

Geometry 2)
We could be observing emission from the shielded face of NLR clouds on the same side of the source as our line of sight. The {\sc Cloudy} outputs of transmitted lines and diffuse emission were additively combined to approximate this geometry. The best fit parameters for this geometry gave a C-stat value of $\sim$300 larger than the best fit for geometry 1), with smaller $N_H$, but very similar $\xi$ and $v_{turb}$ values ($\log \xi$ $=$ 1.5, $\log N_H$ $=$ 22 and $\log v_{turb}$ $=$ 2.5).

Geometry 3)
We could be observing a combination of the two above geometries: emission from both the illuminated face of NLR clouds on the far side of the source and the shielded face of NLR clouds on the near side of the source, with respect to our line of sight. To model this we used a combination of the outputs used for geometries 1) and 2) above.
To restrict the permutations of this geometry, and therefore reduce computational time, we made the simplifying assumption that both the near and far NLR clouds have the same parameters ($\xi$, $N_H$ and $v_{turb}$).
The reflected output (representing emission from clouds on the far side of the source) was given a positive velocity shift (+300\,\kms) and the transmitted outputs (representing emission from clouds on the near side of the source) were given a negative velocity shift of (-300\,\kms), so that both are outflowing away from the source.
We tested this geometry in three ways: i) with no absorption of either emission components, ii) with absorption of the far clouds' emission (the reflected output) by WA components B and E and no absorption of the near clouds' emission, and iii) with absorption of the far clouds' emission (the reflected output) by all WA components (A-F) and no absorption of the near clouds' emission.
The first form of this geometry, i), also represents the two-cone scenario we discussed in Sect. \ref{2cone_sect}; while our previous tests using {\sc spex} indicated this geometry is not applicable to {\object NGC 5548}, we have included it here for completeness within our {\sc Cloudy} tests.
In all three cases, i), ii) and iii), the best fit parameters were: $\log \xi$ $=$ 1.5, $\log N_H$ $=$ 22.25 and $\log v_{turb}$ $=$ 2.5. The best of these cases, in terms of C-statistic value, is ii), but even in this case the C-statistic value is $\sim$200 larger (d.o.f. $=$ 1083) than the best fit for geometry 1).

These tests have shown that, when comparing these three simple geometries, emission from the illuminated face of the cloud modelled in {\sc Cloudy} is the best match to the narrow emission line data of {\object NGC 5548}. The simplest interpretation is that the NLR we see is emission from the illuminated face of clouds on the far side of the source. Assuming there are absorbing clouds (WAs) closer to the source than the NLR (or on the near side of the source), then absorption of the NLR emission by these clouds would explain the velocity discrepancy seen in the O VII triplet (see Sect \ref{abs_sect}).
The problem with this interpretation is that these emission lines are blueshifted, therefore if this emission was coming from the far side of the source the emitting material would be inflowing towards the source, not outflowing from it, as we would expect.

It is very likely that the geometry in this system is more complex than the plane-parallel cloud assumed in {\sc Cloudy} to cause this, so we cannot be more specific at this stage.

%
\section{Discussion}
\label{discussion_sect}

As previously stated, the 2013-14 770\,ks RGS spectrum shows the largest number of soft X-ray narrow emission lines and RRC features ever observed in \object{NGC 5548}. In total, fourteen narrow lines and eight RRCs are detected, with upper limits for one more narrow emission line. Due to the low soft X-ray continuum flux during these observations, we identify six narrow lines and six RRCs which had not previously been detected in \object{NGC 5548}. We also present new parameters for eight narrow emission lines and two RRCs identified in previous observations of this source.

We detect six of the ten narrow lines found by \cite{Steenbrugge:2005dr}; one of their lines (C V f) falls outside our wavelength range, and the other three (Mg XI f, Al XII f \& Si XIII f) are in the steep, noisy short wavelength part of our spectrum, so it is very difficult to set stringent limits to their parameters using these data.

Using photoionisation models from {\sc Cloudy} we have constrained the emitting gas to an ionisation parameter of log $\xi$ $= 1.45 \pm 0.05$, column density of log $N_H$ $= 22.9 \pm 0.4$\,cm$^{-2}$ and turbulent velocity of log $v_{turb}$ $= 2.25 \pm 0.5\,$\kms.
Using the same conversion between the ionisation parameters of $\xi$ and U as in \cite{Arav:2014ip} for the historical SED of \object{NGC 5548} ($\log\,\xi=\log\,$U$ + 1.6$), we gain an estimate for our emitting gas of $\log\,$U$ =-0.15\pm0.05$. We can now compare both the ionisation parameter and column density from this work to similar work on other Seyfert objects (see Table \ref{otherauthors_table} for a summary of previous work).
The column density value we find (10$^{22.9\pm0.4}$\,cm$^{-2}$) is similar to the highest values found for other objects. For example \object{NGC 4051}, \object{NGC 4151} and \object{NGC 1365} \citep[by][respectively]{Nucita:2010fn,Armentrout:2007bh,Guainazzi:2009fv} are all found to need emitting gas with column densities of the order of $10^{23}$\,cm$^{-2}$ in order to model their X-ray NLR emission.
Our NGC 5548 ionisation parameter, converted to $\log$\,U$ =-0.15\pm0.05$, is lower than most for other Seyfert galaxies, except for the low ionisation component of the three phase model for \object{NGC 4151} \citep{Armentrout:2007bh}.
As expected, our ionisation parameter is higher than that needed to model the UV narrow emission lines in \object{NGC 5548} itself \citep{Kraemer:1998he}.
Due to the much higher soft X-ray continuum flux in previous observations of \object{NGC 5548}, and therefore a much smaller number of measurable X-ray narrow emission lines, this is the first time photoionisation modelling has been used to constrain the ionisation parameter and column density of the X-ray narrow-line emitting gas in this galaxy.

\subsection{Warm absorption of the NLR}
\label{OVII_outflows_sect}

Our initial fit (using a \gaussian model in {\sc spex} for each narrow line) to the 2013-14 stacked spectrum of 770\,ks shows a difference in velocity between the O VII resonance line ($-20 \pm 50\,$\kms) and the forbidden and intercombination lines ($-320 \pm 40\,$\kms and $-480 \pm 160\,$\kms, respectively). This result is hard to interpret physically; the triplet lines should all be formed by the same plasma and therefore have consistent velocities. We investigated possible explanations for this mismatch of velocities such as calibration effects,
emission from two ionisation cones, and absorption, concluding that absorption is most likely (as detailed in Sect. \ref{triplet_sect}).

An alternative hypothesis could be considered, that the O VII triplet may be a superposition of emission by photoionised and collisional plasmas, such as found in star forming regions. However, this is very unlikely; \object{NGC 5548} does not have a high star formation rate, and the AGN is too bright to allow any star formation rate contribution to be detected unless it were a very significant one \citep{1998AstL...24..160D}.

Our main result, which resolves the velocity discrepancy of the O VII triplet, is by having the narrow emission lines absorbed by at least one of the six X-ray WA components found by \cite{Kaastra:2014ip}.
The implications of this in terms of the distance of the NLR from the source are discussed in Sect. \ref{distance_sect} below.

A similar result has been obtained for the UV narrow emission lines in \object{NGC 7469}, where \cite{Kriss:2003ft} found that the best fit to their Far Ultraviolet Spectroscopic Explorer (\Fuse) data included absorption of the narrow emission line components by the UV warm absorbers, at the same covering fraction as the absorption of the broad emission lines and continuum.

Using the best fit model from the scenario with absorption of both the narrow emission lines and RRC features by WA components B and E, we calculated the ratio of the unabsorbed flux of O VII RRC to the O VII forbidden line (RRC/f), as was done in \cite{Kaastra:2002ic}.
The 2013-14 absorption scenario RRC/f value is $0.44 \pm 0.05$, much closer to (and consistent within 3$\sigma$ to) the value of 0.57 from \cite{2002ASPC..255...43B} photoionisation models for an adopted temperature of {5$\,$eV \footnote{As the temperature dependence of this ratio is weak \citep{Kaastra:2002ic}, and our best fit temperature from the pair of Oxygen RRCs (O VII and O VIII) is $6.6 \pm 0.06\,$eV, this is a reasonable comparison to make}} than the value from the 1999 observations \citep[$0.07 \pm 0.12$;][]{Kaastra:2002ic}.
Within \cite{Kaastra:2002ic} the authors suggest that their smaller than expected RRC/f value may be due to a combination of limited statistics and nearby WA absorption features contaminating the RRC flux measurement.
Without WA absorption of either narrow emission lines or RRCs, a smaller than expected  RRC/f value is also observed in the 2013-14 data ($0.28\pm0.03$), which again is not at all consistent with the model value.
Therefore including the absorption by WA components B and E of both the RRCs and the narrow emission lines greatly reduces the tension between observed and modelled O VII RRC/f values, further supporting this absorption scenario.

\subsection{NLR reacting to unobscured SED}
\label{OVII_variability_sect}

Our work using {\sc Cloudy} (see Sect. \ref{cloudy_sect}), with both RGS and EPIC data, shows that the narrow-line emitting gas is still reacting to the unobscured SED.
This implies that we should not expect any additional variation in the narrow lines compared to previous observations and levels of variability between them.
As previous analysis of historical observations did not take into account any absorption of the narrow lines by the WAs that we now see, in order to make appropriate comparisons we must make the same assumptions as in previous analyses.
The 2013-14 O VII f unabsorbed flux, calculated without including any absorption by WA components, is 0.65$^{+0.01}_{-0.05}$\,photons\,m$^{-2}$\,s$^{-1}$, which is between values from observations between 1999 and 2001 (4 observations, with flux values $0.81\pm0.16$, $0.82\pm0.18$, $1.3\pm0.2$ and $1.1\pm0.1$\,photons\,m$^{-2}$\,s$^{-1}$, in chronological order), and the 2005 and 2007 values ($0.35\pm0.06$ and $0.27\pm0.06$\,photons\,m$^{-2}$\,s$^{-1}$ respectively), when the source was seen in an unobscured but intrinsically low-flux state (these historical measurements of O VII f fluxes are collated in \citealp{Detmers:2009bq}).
The unabsorbed flux of the O VII f line in 2002 (0.75 $\pm$ 0.07\,photons\,m$^{-2}$\,s$^{-1}$) is the historical value closest to our 2013-14 measurement, still differing from it by just over the 1$\sigma$ level.
We conclude that there is marginal evidence for variability between 2013-14 and some previous epochs, which is well within the expected range for this source \citep{Detmers:2009bq}. This supports our claim that the NLR is still reacting to the unobscured SED, as it would have been during all these previous epochs.


\subsection{Distance estimates of \object{NGC 5548}'s X-ray NLR}
\label{distance_sect}

In our best fit absorption scenario, the emitting gas (determined from {\sc Cloudy} modelling) has the parameters: log $\xi$ $= 1.45 \pm 0.05$, log $N_H = 22.9 \pm 0.4$ and log v$_{turb}$ $= 2.25 \pm 0.5$.

Using this result, and the definition of $\xi$ ($\xi \equiv L/nr^{2} $, Sect. \ref{intro_sect}) we can determine a lower limit to the distance of the emitting gas from the source. We use $n < 5 \times 10^{9}$\,cm$^{-3}$ as this is the critical density for O VII \citep{Porquet:2000cv} and as we know this gas is photoionised (from the presence of narrow RRCs), the density must be below this value. $L$ ($= 11.7 \times 10^{43}$\,ergs\,s$^{-1}$) is taken from \cite{Mehdipour:2014ip}, and is the luminosity of the unobscured 2013-14 SED, and $\xi$ (log $\xi$ $=$ 1.45) is taken from the {\sc Cloudy} best fit to the data. This gives a (not very restrictive) lower limit to the distance of the emitting gas of $r > 9.3 \times 10^{-3}$\,pc.

For another distance estimate, we can begin by writing the column density, $N_H$, as:
\begin{equation}
N_H = \int_{r_{min}}^{r_{max}} n_H dr = \frac{n_H}{n_e} \frac{L}{\xi} (\frac{1}{r_{min}} - \frac{1}{r_{max}})
\end{equation}
which, using the same method as \cite{Behar:2003ho} and assuming $r_{max} \gg r_{min}$, can be rewritten as:
\begin{equation}
r_{min} \simeq (r_{min}^{-1} - r_{max}^{-1})^{-1} = \frac{n_H}{n_e} \frac{L}{\xi N_H}.
\end{equation}
We can then use the simplifying relation $n_e = 1.2n_H$ (for a fully ionized cosmic plasma), and our values of $L$, $\xi$ and $N_H$ (see previous paragraph) to calculate $r_{min} = 13.9\pm0.6$\,pc. This is consistent with the previous distance estimate of 1-15\,pc for the X-ray NLR by \cite{Detmers:2009bq}, especially when considering that the {\sc Cloudy} modelling in this work suggests that the narrow-line emission comes mainly from the illuminated face of the NLR cloud(s), which would be located at $r_{min}$.

We note that the log $\xi$ value of the emitting gas found from {\sc Cloudy} analysis, 1.45 $\pm$ 0.05, is consistent with the 2002 $\log \xi$ value of WA component B (1.51 $\pm$ 0.05, when the WAs were reacting to the unobscured ionizing flux). Combining this with the consistent distance estimates for the X-ray NLR ($r_{min}=13.9\pm0.6$\,\,pc, this work; 1-15\,pc, \citealp{Detmers:2009bq}) and WA component B ($<$40\,pc, \citealp{Ebrero:2014ip}), can we identify the NLR emission gas with the WA absorption gas? The column density for the emitting gas determined with {\sc Cloudy}, log $N_H$ $= 22.9\pm0.4\,$cm$^{-2}$, is higher than the combined column densities of all WA components. However, WAs are seen along the line of sight while the X-ray NLR emission is likely coming from a much larger region, given the fact that it is still reacting to the unobscured SED, so there could be column density variations over that region. The coincidence of ionisation parameters and compatible distances of the X-ray NLR gas with one of the WA components is not seen here as enough evidence to conclusively identify any of the WA components as the same physical gas as the narrow-line emitting gas, although it is suggestive of that possibility.

\section{Conclusions}
\label{conclusions}

We have used the 770\,ks RGS spectrum of \object{NGC 5548} from our 2013-14 observational campaign to show evidence that the narrow emission lines undergo absorption by at least one of the six known warm absorber (WA) components in this source. There are indications that the main absorption comes from WA component B (labelled as such by \citealp{Kaastra:2014ip}). We have established this from careful analysis of the O VII triplet, in the wavelength range 21.5-23$\,\AA$. This solution also resolves tension between modelled and observed values of the O VII RRC/f unabsorbed flux ratio.
Unfortunately the data do not allow us to conclusively distinguish between competing detailed scenarios of absorption of the narrow emission lines by different combinations of WA components. The presence of this absorption is a more important result than which of the specific components are causing it.

Through comparison to simulated spectra produced by the photoionisation code {\sc Cloudy}, we use both RGS and EPIC-pn data to determine that the emitting gas can be described as reacting to the unobscured SED (i.e. to ionising radiation before the intervention of the `obscurer' discovered in our observing campaign). The emitting gas has a well constrained ionisation parameter of log $\xi$ $= 1.45 \pm 0.05$ and column density and turbulent velocity of log $N_H$ $= 22.9 \pm 0.4$\,cm$^{-2}$ and log $v_{turb}$ $= 2.25 \pm 0.5\,$\kms respectively. This is the first time the X-ray NLR emitting gas in this source has been characterised by an ionisation parameter and column density. From this we estimate the distance of these emitting clouds from the central source as $13.9 \pm 0.6$\,pc.

\begin{acknowledgements}
The data used in this research are stored in the public archives of the satellites that are involved. We thank the International Space Science Institute (ISSI) in Bern for support. This work is based on observations obtained with \xmm, an ESA science mission with instruments and contributions directly funded by ESA Member States and the USA (NASA). M. W. acknowledges the support of a PhD studentship awarded by the UK Science \& Technology Facilities Council (STFC). SRON is supported financially by NWO, the Netherlands Organization for Scientific Research. M.M. acknowledges the support of a Studentship Enhancement Programme awarded by the UK Science \& Technology Facilities Council (STFC). K.C.S. acknowledges financial support from the Fondo Fortalecimiento de la Productividad Cient\'\i fica VRIDT 2013. E.B. is supported by grants from Israel's MoST,  ISF (1163/10), and I-CORE program (1937/12). B.M.P. acknowledges support from the US NSF through grant AST-1008882. M.C. and S.B. acknowledge INAF/PICS support. G.P. acknowledges support via an EU Marie Curie Intra-European fellowship under contract no. FP-PEOPLE-2012-IEF-331095. P.O.P acknowledges funding support from the CNES and the French-Italian International Project of Scientific Collaboration: PICS-INAF project n181542. This research has made use of the NASA/IPAC Extragalactic Database (NED), which is operated by the Jet Propulsion Laboratory, California Institute of Technology, under contract with the National Aeronautics and Space Administration. This research has made use of NASA's Astrophysics Data System Bibliographic Services'. M.W. thanks Mat Page for useful discussions. The authors thank the anonymous referee for their useful suggestions.

\end{acknowledgements}

\bibliography{BibliographyMay14}

\end{document}